\documentclass[12pt, twoside]{article}
\usepackage{amsmath,amssymb}
\usepackage{float}
\usepackage{multirow}
\usepackage{graphicx}
\usepackage{epsfig}
\usepackage{cite}
\usepackage{cases}
\usepackage{epstopdf}
\usepackage{caption}
\usepackage{titlesec}
\usepackage{textcomp}
\usepackage{textcomp,booktabs}
\usepackage[usenames,dvipsnames]{color}
\usepackage{colortbl}
\definecolor{mygray}{gray}{0.9}
\definecolor{mypink}{rgb}{0.99,0.91,0.95}
\definecolor{mycyan}{cmyk}{0.3,0,0,0}
\usepackage{booktabs}
\usepackage[top=2.5cm,bottom=2.5cm,left=2cm,right=2cm]{geometry}
\usepackage[colorlinks,linkcolor=red,anchorcolor=green,citecolor=blue,CJKbookmarks=True]{hyperref}

\allowdisplaybreaks
\usepackage{diagbox}

\begin{document}
\title{\textbf{An Improved Inverse Method for Estimating Disease Transmission Rates in Low-Prevalence Epidemics}}
\author{Shuanglin Jing$^{a,b}$, Yuting Huang$^{c}$, Hai-Feng Huo$^{a,b}$\footnotemark[1]\\
$^{a}$School of Mathematics and Physics, Lanzhou Jiaotong University, \\Lanzhou, Gansu, 730070, China\\
$^{b}$Gansu Center for Fundamental Research in Complex Systems \\
Analysis and Control, Lanzhou Jiaotong University, Lanzhou, \\
Gansu, 730070, China\\
$^{c}$The Third People's Hospital of Lanzhou, Lanzhou, Gansu, 730050, China}
\renewcommand{\thefootnote}{\fnsymbol{footnote}}
\footnotetext[1]{Corresponding author. Email: hfhuo@lut.edu.cn}
\date{}
\maketitle
\begin{abstract}
The accurate estimation of time-varying transmission rates is fundamental for understanding infectious disease dynamics and implementing effective public health interventions. To this end, we propose an improved inverse method for estimating time-varying transmission rates in low-prevalence settings, where conventional data preprocessing approaches often fail due to sparse case observations. To overcome this difficulty, we introduce an exponential B-spline interpolation approach that integrates both continuous and discrete inverse methods. This method ensures that transmission rate estimates remain non-negative and smooth, even when the observed data exhibit low cases. We apply this approach to several infectious disease models using real-world data from China, including a scarlet fever model, a multi-strain influenza model, and an age-structured influenza model. The results show that our method provides accurate transmission rate estimates, particularly in low-prevalence infectious diseases and multi-group epidemic models, demonstrating its robustness and applicability across various epidemiological contexts. The improved inverse method offers a new perspective for epidemiological modeling and provides reliable technical support for related theoretical exploration and public health decision-making.

\end{abstract}

\section{Introduction}
The transmission rate of infectious diseases, particularly time-varying transmission rates, is a core parameter in characterizing the dynamical features of epidemics \cite{wang2023discrete}. Accurate estimation of this parameter is crucial for formulating targeted prevention strategies and evaluating intervention effectiveness \cite{lange2016reconstruction}. In the global public health emergency system, low-prevalence diseases, such as newly emerging infectious diseases sporadically occurring in specific regions or periodically recurring low-prevalence diseases, may not cause large-scale outbreaks. However, their latent transmission risks, potential for cross-regional spread, and long-term public health burden cannot be ignored \cite{wang2023discrete}. The transmission process of these diseases has distinct characteristics: the infection cases are sparse and geographically dispersed, with long incubation periods and hidden transmission chains. Traditional case surveillance data often suffer from delays or incompleteness, making the estimation of transmission rates particularly challenging \cite{wang2023discrete}. For example, in small closed populations such as schools or ships, the spread of epidemics in low-prevalence settings is often difficult to capture accurately due to insufficient sample size and data noise interference, necessitating the development of targeted modeling and estimation methods.

In the field of epidemiological modeling, inverse methods, which couple observational data with dynamic models to reverse-engineer key unknown parameters, have become one of the mainstream techniques for estimating core parameters such as transmission rates \cite{mummert2019parameter,jing2025infectivity,kong2015inverse,jagan2020fast,wang2023discrete,lange2016reconstruction}. Current inverse methods for time-varying transmission rate estimation can be categorized into three types: The first category consists of algorithms based on probabilistic inference, such as Markov Chain Monte Carlo (MCMC) \cite{haario2006dram}, sequential Monte Carlo (SMC) \cite{doucet2001sequential}, and Approximate Bayesian Computation Sequential Monte Carlo (ABC-SMC) \cite{toni2009approximate,marin2012approximate}. These methods typically parameterize the transmission rate as a quasi-smooth function, such as Spline functions, Fourier series, or Legendre polynomials, and reconstruct the time-varying transmission rate by estimating the node parameters of the spline functions or the coefficients of the Fourier series and Legendre polynomials \cite{jing2025infectivity,xue2022infectivity,he2023resolving,XueEvaluating2022,cazelles2018accounting,smirnova2019forecasting}. The second category includes Physics-Informed Neural Networks (PINNs) \cite{raissi2019physics,LuDeepXDE2021}, where the transmission rate is modeled as a neural network function, and parameter estimation is achieved by combining data-driven approaches with physical constraints \cite{wang2025discovering,tang2024managing,he2023combining,song2022estimating}. The third category is direct inverse methods, which do not rely on machine learning algorithms but instead directly solve for unknown parameters through rigorous mathematical derivation \cite{kong2015inverse,jagan2020fast,wang2023discrete,wang2025estimation}. Notably, continuous inverse methods proposed by Kong et al.\cite{kong2015inverse} and Pollicott et al. \cite{pollicott2012extracting}, as well as the discrete inverse methods introduced by Wang and Wang \cite{wang2022hypothesis,wang2022policy,wang2023discrete}, have demonstrated significant practical value in real-world applications \cite{ji2023climate,ji2025hybrid}. Both continuous and discrete inverse methods do not rely on machine learning algorithms or statistical inference techniques, offering the advantage of computational efficiency, which can save considerable time. Additionally, the local lagged adapted generalized method of moments (LLGMM) \cite{mummert2019parameter}, a parameter identification method for stochastic infectious disease models, and a directly derived method for estimating time-varying recovery and mortality rates \cite{GhoshEstimation2025} provide important supplements to related research.

However, despite the prominent practical value of the aforementioned direct inverse methods proposed by Kong et al. \cite{kong2015inverse}, Pollicott et al. \cite{pollicott2012extracting}, Wang and Wang \cite{wang2023discrete}, the authors face a common problem when applied to low-prevalence infectious diseases. Specifically, when new case counts remain at a low-prevalence for an extended period or even approach zero, traditional data preprocessing or interpolation techniques (such as polynomial interpolation) are prone to generating negative estimates, which in turn lead to distorted transmission rate calculations and even method failure. This limitation greatly restricts the application of such direct inverse methods in scenarios such as the early stages of epidemics, post-effective interventions, or surveillance of endemic low-prevalence diseases.

To address this problem, we propose an exponential B-spline interpolation method and integrate it into both continuous and discrete inverse methods frameworks. Specifically, we first take the logarithm of new cases, then interpolate the logarithmic values using B-splines \cite{prautzsch2002bezier}, and finally exponentiate the interpolated logarithmic new cases. In this way, we construct a positive-preserving and smooth interpolation curve that effectively avoids the generation of negative values, significantly enhancing the robustness and reliability of transmission rate estimation in low-prevalence scenarios. In this work, we systematically present this improved inverse method, demonstrating its superior performance in handling low-prevalence epidemic data. We apply it to models of low-prevalence endemic childhood diseases, multi-strain infectious diseases, and age-structured infectious disease models, verifying its broad applicability.

The rest of the work is organized as follows. In Section \ref{section2}, we propose a non-autonomous SEIR model and prove the non-negativity of its solutions. Subsequently, we introduce exponential B-spline interpolation and re-derive the process of estimating the transmission rate using both continuous and discrete inverse methods. Finally, we compare the improved inverse method with the original one via illustrative examples. In Section \ref{section3}, the improved inverse method is applied to scarlet fever data from Jiangxi Province and the Tibet Autonomous Region of China to verify its effectiveness for low-prevalence infectious diseases. In Section \ref{section4}, we focus on multi-strain influenza transmission in China, where we derive the expression for time-varying transmission rates in a multi-strain model and validate it using pre- and post-COVID-19 pandemic data. In Section \ref{section5}, we conduct inverse method derivation for an age-structured influenza model and test its performance across annual cycles and low-incidence periods using historical data from China. In Section \ref{section6}, we summarize the advantages, potential limitations, and future directions of the proposed inverse method.

\section{Time-varying transmission rates of SEIR model}\label{section2}
To estimate the time-varying transmission rate of the infectious disease model, the total population at time $t$, denoted as $N(t)$, is divided into four compartments: $S(t), E(t), I(t)$, and $R(t)$, such that $N(t)=S(t)+E(t)+I(t)+R(t)$. Here, $S(t)$ represents susceptible individuals, $E(t)$ denotes exposed  individuals, $I(t)$ denotes the infected individuals, and $R(t)$ denotes individuals that have recovered and are fully immune.  The model expression is given by
\begin{eqnarray}
&&\;\frac{\mbox{d}S(t)}{\mbox{d}t}=\Lambda-\frac{\beta(t)S(t)I(t)}{N(t)}-dS(t),\nonumber\\
&&\left.
\begin{split}
&\frac{\mbox{d}E(t)}{\mbox{d}t}=\frac{\beta(t)S(t)I(t)}{N(t)}-(\sigma+d)E(t),\\
&\frac{\mbox{d}I(t)}{\mbox{d}t}=\sigma E(t)-(\gamma+d)I(t),\\
\end{split}
\right.\label{EQ1}\\
&&\;\frac{\mbox{d}R(t)}{\mbox{d}t}=\gamma I(t)-d R(t),\nonumber
\end{eqnarray}
where $\Lambda$ represents the recruitment rate of new individuals into the susceptible compartment, $\sigma$ denotes the rate of progression from exposed to infected individuals, $\gamma$ is the recovery rate of infected individuals, $d$ indicates the natural death rate, and $\beta(t)$ is the time-varying transmission rate.

To prove the nonnegativity of the Model \(\eqref{EQ1}\), we need to show that if the initial values satisfy $S(0) > 0$, $E(0) > 0$, $I(0) > 0$, $R(0) > 0$, then $S(t) > 0$, $E(t) > 0$, $I(t) > 0$, $R(t) > 0$ for all $t > 0$. We define
$$W(t) = \min\Big\{S(t), E(t), I(t), R(t)\Big\},\;\text{for}\;\;t > 0.$$
It is clear that $W(0) > 0$. Suppose there exists $t_1 > 0$ such that $W(t_1) = 0$ and $W(t) > 0$ for all $t \in [0, t_1)$.

We only consider the case where $W(t_1) = S(t_1)$. For $t \in [0, t_1]$, we have $E(t) \geq 0$, $I(t) \geq 0$, and $R(t) \geq 0$, and the total population \( N(t) = S(t) + E(t) + I(t) + R(t) \geq S(t) > 0 \). Since $I(t) \geq 0$, $\beta(t) \geq 0$, $d > 0$, and $\Lambda>0$, we have
$$
\frac{{\rm d}S}{{\rm d}t} \geq - S(t)\left( \frac{\beta(t)I(t)}{N(t)} + d \right).
$$
By the integrating factor method, we obtain
$$
S(t) \geq S(0) \exp\left( - \int_0^t \left( \frac{\beta(\tau)I(\tau)}{N(\tau)} + d \right) {\rm d}\tau \right).
$$
When $t = t_1$, $S(t_1) = 0$, but the right-hand side is $0$ multiplied by a positive exponential function, which is a contradiction. Thus, $S(t) > 0$ for all $t \geq 0$.

Similarly, we can also prove that $E(t) > 0$, $I(t) > 0$, and $R(t) > 0$ for all $t \geq 0$.

\subsection{Improved inverse method}\label{subsection2.1}
\begin{figure}[h]
\centering
{\includegraphics[width=5in,height=3.5in,clip]{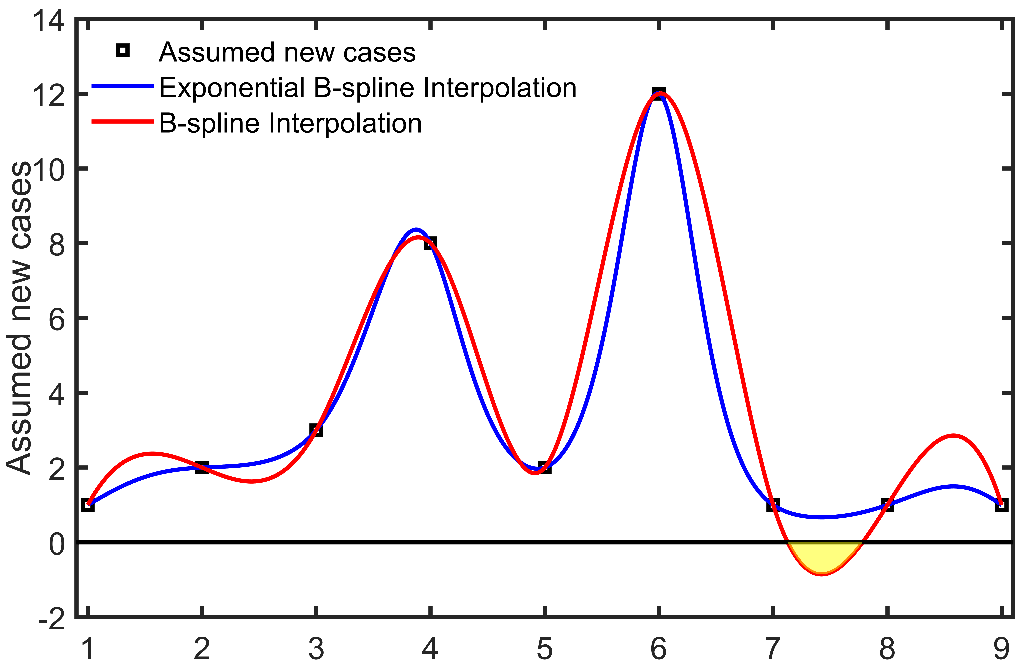} }
\renewcommand{\figurename}{\textbf{Figure.}}
\caption{Comparison between exponential B-spline interpolation and B-spline interpolation.}\label{fig1}
\end{figure}

In this subsection, we introduce improved continuous and discrete inverse methods to accurately estimate the time-varying transmission rate \(\beta(t)\) in Model (\ref{EQ1}) from new cases. We first denote the number of newly observed cases on day (week, month, etc.) $j$ as
$$y_j>0,\quad j=0,1,\dots,M.$$
To address the mismatch between the observed time intervals of $y_j$ and the time step of the model, data interpolation is necessary. It should be noted that in low-prevalence epidemic contexts, such direct interpolation may introduce numerical artifacts in the form of negative values, as shown in Figure \ref{fig1}. Thus, we take the logarithm of $y_j$ to obtain
$$Y_{j}=\ln(y_{j}), \quad j = 0, 1, \dots, M.$$
We then select the B-spline basis functions $\{B_{i,p}(t)\}_{i=1}^m$ with given order and knots, and construct the B-spline interpolation function \cite{prautzsch2002bezier} for the logarithmic number of cases on the interval $[0, M]$:
\begin{equation}
Y^{\mathrm{BSI}}(t) = \sum_{i=1}^m P_i B_{i,p}(t), \quad t \in [0, M],\label{EQ2}
\end{equation}
where $B_{i,p}(t)$ denotes the $p$-th degree B-spline basis function, $P_i$ are the B-spline coefficients, $m$ is the number of control points, and $p$ is the degree of the spline. Note that the coefficients $P_i$ are uniquely determined by the interpolation conditions:
$$
Y^{\mathrm{BSI}}(j) = Y_{j}=\ln(y_{j}), \quad j = 0, 1, \dots, M.
$$
Next, we derive the transmission rate through both the continuous inverse method and the discrete inverse method.
\subsubsection{Improved continuous inverse method}
We begin by applying the exponential function to both sides of Eq. (\ref{EQ2}), which yields
\begin{equation}
\tilde{y}(t) = \exp\left(Y^{\mathrm{BSI}}(t)\right) = \exp\left(\sum_{i=1}^m P_i B_{i,p}(t)\right), \quad t \in [0, M].\label{EQ3}
\end{equation}
In Model (\ref{EQ1}), $\sigma E(t)$ denotes the number of new cases per unit time, which is consistent with the biological interpretation of $\tilde{y}(t)$. Thus, we have
$$\sigma E(t)=\tilde{y}(t).$$
We let $p \geq 2$, then $Y^{\mathrm{BSI}}(t)$ admits a first-order derivative. Accordingly, we have
$$\frac{\mbox{d}E(t)}{\mbox{d}t}=\frac{1}{\sigma}\frac{\mbox{d}\tilde{y}(t)}{\mbox{d}t}=\frac{1}{\sigma}\exp\left(Y^{\mathrm{BSI}}(t)\right)\frac{\mbox{d}Y^{\mathrm{BSI}}(t)}{\mbox{d}t}.$$
For convenience, we set $ H(t):=\frac{{\rm d}E(t)}{{\rm d}t}$.

Next, we solve for \( N(t) \), \( S(t) \), \( I(t) \), and \( R(t) \). Given the ordinary differential equation describing the dynamics of a population \(N(t)\):
\[
\frac{{\rm d}N(t)}{{\rm d}t} = \Lambda - d N(t), \quad N(0) = N_0.
\]
Its solution is given by
\[
N(t) = N_0 {\rm e}^{-d t} + \frac{\Lambda}{d} \left(1 - {\rm e}^{-d t}\right).
\]
The solution to this first-order linear non-homogeneous differential equation for \(I(t)\) takes the form
$$
I(t) = {\rm e}^{-(\gamma + d) t} \left( I_0 +\int_0^t {\rm e}^{(\gamma + d) s} \tilde{y}(s) \, {\rm d}s \right),
$$
where $I_0=I(0)$ denotes the initial condition. Similarly, the differential equation for \(R(t)\) is also a first-order linear non-homogeneous equation. Its solution is given by
$$
R(t) = {\rm e}^{-d t} \left( R_0 + \gamma \int_0^t {\rm e}^{d s} I(s) \, {\rm d}s \right),
$$
where $R_0=R(0)$, and \(I(s)\) denotes the solution for the infectious compartment at time \(s\). The variable $S(t)$ can be obtained as
$$
S(t)=N(t)-\frac{\tilde{y}(t)}{\sigma}-I(t)-R(t)
$$
through $N(t)$, $I(t)$, $R(t)$, and $\tilde{y}(t)$.

Based on the nonnegativity of Model (\ref{EQ1}), the following can be derived from its second equation:
\begin{equation*}
\beta(t)=\frac{\Big[H(t)+(\sigma+d)\frac{\tilde{y}(t)}{\sigma}\Big]N(t)}{S(t)I(t)}, \quad t \in [0, M].
\end{equation*}
To ensure that $\beta(t)\geq0$, we set
\begin{equation}
\beta(t)=\max\Bigg\{0,\frac{\Big[H(t)+(\sigma+d)\frac{\tilde{y}(t)}{\sigma}\Big]N(t)}{S(t)I(t)}\Bigg\}, \quad t \in [0, M].\label{EQ4}
\end{equation}

\subsubsection{Improved discrete inverse method}
Wang et al. discretized the model using the Forward Euler and Backward Euler methods \cite{wang2023discrete}. They found that the Forward Euler method offers higher accuracy and faster speed. Therefore, the discrete SEIR Model (\ref{EQ1}) based on the Forward Euler discretization method can be expressed as
\begin{eqnarray}
&&S_{n+1}=S_{n}+\bigg(\Lambda-\frac{\beta_{n}S_{n}I_{n}}{N_{n}}-dS_{n}\bigg)\Delta t,\nonumber\\
&&E_{n+1}=E_{n}+\bigg(\frac{\beta_{n}S_{n}I_{n}}{N_{n}}-(\sigma+d)E_{n}\bigg)\Delta t,\nonumber\\
&&I_{n+1}=I_{n}+\Big(\sigma E_{n}-(\gamma+d)I_{n}\Big)\Delta t,\label{EQ5}\\
&&R_{n+1}=R_{n}+\Big(\gamma I_{n}-d R_{n}\Big)\Delta t,\nonumber\\
&&N_{n+1}=N_{n}+\Big(\Lambda-d N_{n}\Big)\Delta t,\nonumber
\end{eqnarray}
where $n = 0, 1, \dots, K$. Since the model adopts a time step $\Delta t$, we set
$$
t_n = n\Delta t, \quad n = 0, 1, \dots, K.
$$
At these time points, we take the exponential of $Y^{\mathrm{BSI}}(t)$ to yield a sequence of case counts consistent with the time steps of the model:
\begin{equation}
\tilde{y}_n = \exp\left(Y^{\mathrm{BSI}}(t_n)\right) = \exp\left(\sum_{i=1}^m P_i B_{i,p}(t_n)\right), \quad n = 0, 1, \dots, K.\label{EQ6}
\end{equation}
Thus, we obtain the new cases $\tilde{y}_n$ with the same dimension as $\sigma E_n$. Accordingly, we have
\begin{equation}
\sigma E_n = \tilde{y}_n, \quad n = 0, 1, \dots, K.\label{EQ7}
\end{equation}
From Eq. (\ref{EQ7}), we can obtain
$$E_{n}=\frac{\tilde{y}_n}{\sigma}.$$
Thus, $I_{n}$ and $R_{n}$ can be obtained through an iterative process as follows
\begin{eqnarray}
&&I_{n+1}=I_{n}+\Big[\tilde{y}_n-(\gamma+d)I_{n}\Big]\Delta t,\nonumber\\
&&R_{n+1}=R_{n}+\Big(\gamma I_{n}-d R_{n}\Big)\Delta t.\nonumber
\end{eqnarray}
Obviously, $S_{n}$ can be expressed as
$$S_{n}=N_{n}-\frac{\tilde{y}_n}{\sigma}-I_{n}-R_{n}.$$
It follows that
\begin{equation}
\beta_{n}=\max\Bigg\{0,\frac{\Big[(\Lambda-dS_{n})\Delta t-(S_{n+1}-S_{n})\Big]N_{n}}{S_{n}I_{n}\Delta t}\Bigg\}, \quad n = 0, 1, \dots, K-1,\label{EQ8}
\end{equation}
and $\beta_{K}=\beta_{K-1}$.

Figure \ref{fig1} illustrates a comparison between exponential B-spline interpolation and conventional B-spline interpolation in the context of modeling assumed new cases across nine time points. The black square markers denote the assumed new case data points. The blue curve represents the exponential B-spline interpolation, while the red curve corresponds to the B-spline Interpolation. A critical advantage of the exponential B-spline interpolation is evident: the B-spline interpolation (red curve) generates negative values of assumed new cases (e.g., around time point seven), which is biologically implausible for case counts and would induce failures in solving transmission rates. In contrast, the exponential B-spline interpolation (blue curve) sustains non-negative values throughout the entire interval, thus ensuring feasibility in epidemiological modeling and enabling accurate computation of transmission rates.
\subsection{Comparison of improved discrete inverse method and the original discrete inverse method}
\begin{figure}[ht]
\centering
{\includegraphics[width=5.5in,height=4in,clip]{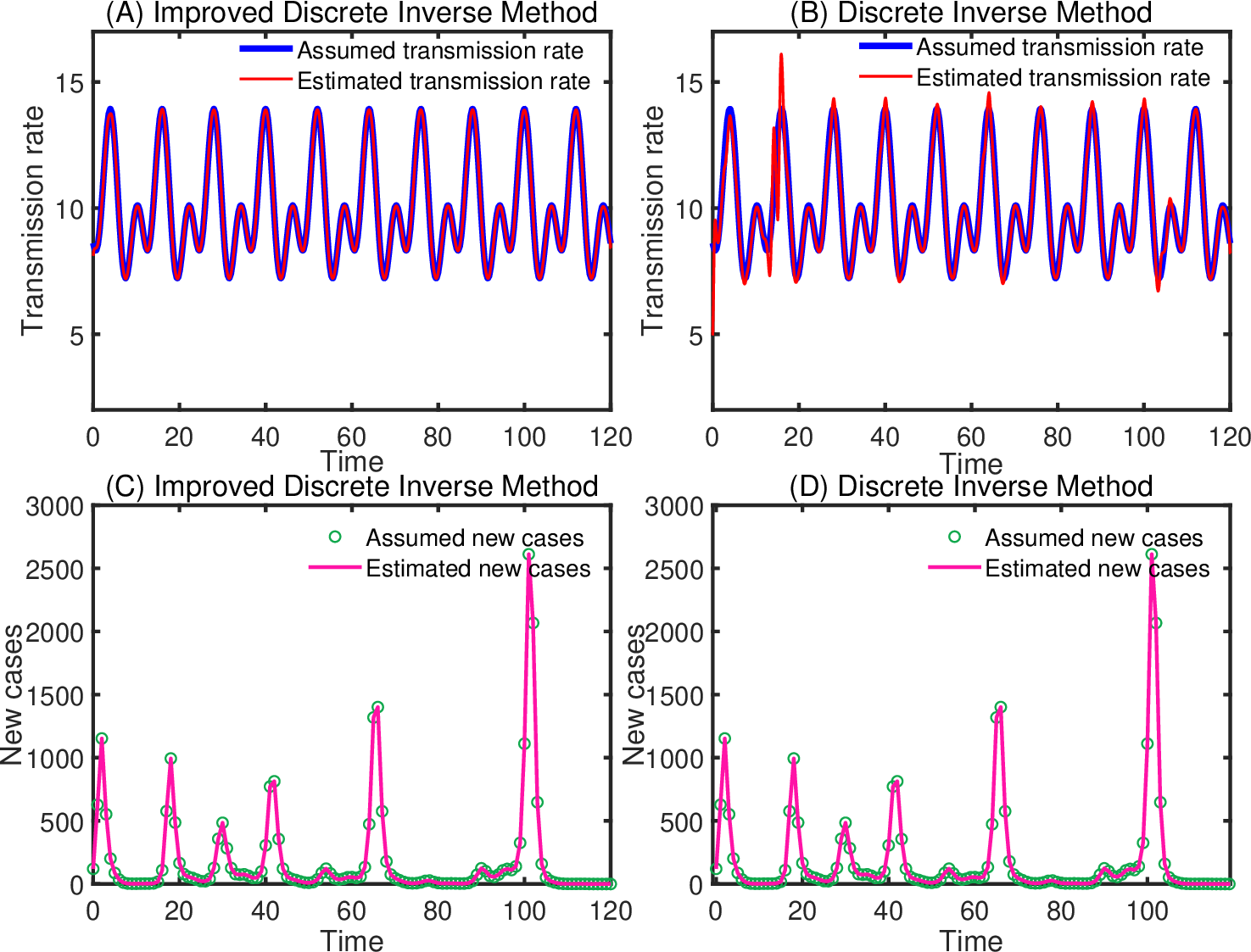} }
\caption{Comparison of the improved discrete inverse method and the original discrete inverse method. Panels (A) and (C) show the comparison between the transmission rates inverted by the improved discrete inverse method and the assumed transmission rates, as well as the comparison between the estimated new cases and the assumed new cases, respectively. Panels (B) and (D) show the comparison of the transmission rates inverted by the original discrete inverse method against the assumed transmission rates, and the comparison of the estimated new cases against the assumed new cases, respectively.}\label{fig2}
\end{figure}

In this subsection, we only compare the advantages of the improved discrete inverse method with the discrete inverse method proposed by Wang et al. \cite{wang2022hypothesis,wang2022policy,wang2023discrete} in inverting transmission rates. The parameters adopted for Model (\ref{EQ5}) are specified as follows. The initial conditions are set as $(S_0, E_0, I_0, R_0) = (2000, 20, 20, 0)$, where $S_0$, $E_0$, $I_0$, and $R_0$ denote the initial numbers of susceptible, exposed, infected, and recovered individuals, respectively. The time step is $\Delta t = 0.001$, with a maximum simulation time of $t_{K} = 120$, where $K = 120000$. The transmission rate $\beta_n$ is assumed to be a time-varying parameter given by
$$\beta_n = 5\left[2 + 0.4\sin\left( \frac{2\pi t_{n}}{12} + 6 \right) + 0.4\cos\left( \frac{2\pi t_{n}}{6} + 2 \right) \right],\quad n = 0, 1, \dots, K,$$
where $t_n = n\Delta t$. Other parameters are set as: the rate of progression from exposed to infected individuals $\sigma = 30/5$; the recruitment rate of new individuals into the susceptible compartment $\Lambda = 300$; the recovery rate of infected individuals $\gamma = 30/7$; and the natural death rate $d = 1/(75 \times 12)$. The results comparing the improved and the original discrete inverse methods are shown in Figure \ref{fig2}. In Panel (A), the estimated transmission rate closely aligns with the assumed one throughout the time period. In contrast, Panel (B) shows noticeable deviations between the estimated and assumed transmission rates.

To present the accuracy comparison between the improved discrete inverse method and the original discrete inverse method, we use the Mean Absolute Error (MAE) to calculate the differences between the assumed and estimated transmission rates, and between the assumed and estimated new cases, for both the improved and the original inverse methods, respectively. The formula of MAE is given by
$$
\text{MAE} = \frac{1}{K+1} \sum_{n=0}^{K} |z_n - x_n|,
$$
where $x_n$ denotes the $n$-th component of that vector for either the assumed transmission rate or the new cases, $z_n$ denotes the $n$-th component of the vector representing either the estimated transmission rate or the new cases, and $K+1$ denotes the dimension of the data vector. Table \ref{Table1} shows the MAE values for transmission rate and new cases. Specifically, for the MAE of transmission rate, the improved discrete inverse method attains a value of 0.0224, while the original discrete inverse method obtains 0.1747. For the MAE of new cases, the improved method records 0.3427, whereas the original method records 4.0449. These results clearly demonstrate that the improved discrete inverse method exhibits superior accuracy in estimating both transmission rate and new cases compared to the original discrete inverse method, as reflected by the notably lower MAE values.
\begin{table}[ht]
\centering
\caption{Comparison of the accuracy between the improved discrete inverse method and the original discrete inverse method.}
\begin{tabular}{|l|l|l|l}
\hline
& Improved Discrete Inverse Method & Discrete Inverse Method\\
\hline
MAE (Transmission rate)& 0.0224& 0.1747 \\
\hline
MAE (New cases)& $0.3427$& $4.0449$ \\
\hline
\end{tabular}\label{Table1}
\end{table}

\section{Application to childhood infectious diseases}\label{section3}
\begin{figure}[ht]
\centering
{\includegraphics[width=5.5in,height=4in,clip]{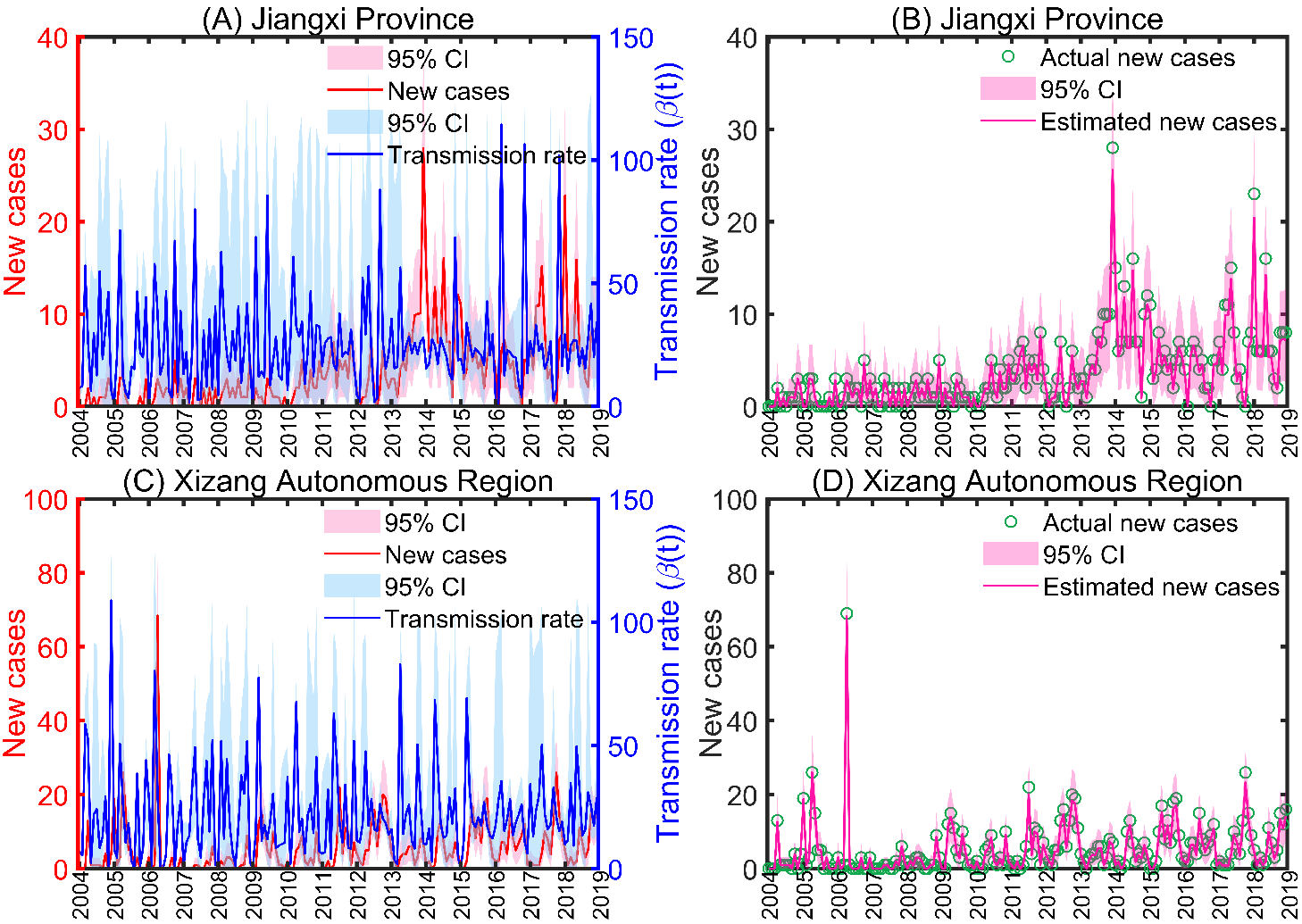} }
\caption{Estimation of the scarlet fever transmission rate. Panels (A) and (B) show the modeling results for Jiangxi Province from 2004 to 2019. (A) shows the temporal variations in both the reported new scarlet fever cases (red line) and the estimated transmission rate $\beta(t)$ (blue line), with their respective 95\% CIs (confidence intervals) shaded in pink and light blue. (B) presents a comparison between the actual new cases (green circles) and the estimated new cases (magenta line with a 95\% CI shaded in pink). Panels (C) and (D) show the modeling results for Xizang Autonomous Region from 2004 to 2019. Specifically, (C) depicts the temporal dynamics of both the reported new scarlet fever cases (red line) and the estimated transmission rate $\beta(t)$ (blue line), while (D) evaluates the model fit by comparing the actual new cases (green circles) against the estimated new cases (magenta line).}\label{fig3}
\end{figure}

Common childhood infectious diseases mainly include varicella, hand, foot and mouth disease (HFMD), measles, mumps, rubella, scarlet fever, and pertussis. These diseases are predominantly transmitted via respiratory droplets, direct contact, or the fecal-oral route. In this section, we establish a childhood infectious disease model and invert the transmission rate using scarlet fever data \cite{liu2020exposure} from Jiangxi Province and the Tibet Autonomous Region of China. Notably, scarlet fever in these two regions is low-prevalence (see Figure \ref{fig3}). Our model \cite{kong2015inverse} is given by
\begin{eqnarray}
&&\frac{\mbox{d}S(t)}{\mbox{d}t}=\Lambda-\frac{\beta(t)S(t)I(t)}{N(t)}-(g+d)S(t),\nonumber\\
&&\frac{\mbox{d}E(t)}{\mbox{d}t}=\frac{\beta(t)S(t)I(t)}{N(t)}-(\sigma+g+d)E(t),\nonumber\\
&&\frac{\mbox{d}I(t)}{\mbox{d}t}=\sigma E(t)-(\gamma+g+d)I(t),\label{EQ9}\\
&&\frac{\mbox{d}R(t)}{\mbox{d}t}=\gamma I(t)-(g+d) R(t),\nonumber\\
&&\frac{\mbox{d}A(t)}{\mbox{d}t}=g(S(t)+E(t)+I(t)+R(t))-d A(t),\nonumber
\end{eqnarray}
where $S$, $E$, $I$, and $R$ represent the susceptible, exposed, infectious, and recovered individuals for the juvenile group, respectively, and $A$ denotes the adult individuals. The parameter $g$ represents the maturation rate of juveniles, while the other parameters carry the same interpretations as those in Model (\ref{EQ1}).

We discretize Model (\ref{EQ9}) using the forward Euler method, and the corresponding discrete model is given by
\begin{eqnarray}
&&S_{n+1}=S_{n}+\bigg(\Lambda-\frac{\beta_{n}S_{n}I_{n}}{N_{n}}-(g+d)S_{n}\bigg)\Delta t,\nonumber\\
&&E_{n+1}=E_{n}+\bigg(\frac{\beta_{n}S_{n}I_{n}}{N_{n}}-(\sigma+g+d)E_{n}\bigg)\Delta t,\nonumber\\
&&I_{n+1}=I_{n}+\Big(\sigma E_{n}-(\gamma+g+d)I_{n}\Big)\Delta t,\label{EQ10}\\
&&R_{n+1}=R_{n}+\Big(\gamma I_{n}-(g+d) R_{n}\Big)\Delta t,\nonumber\\
&&A_{n+1}=A_{n}+\Big[g(S_{n}+E_{n}+I_{n}+R_{n})-d A_{n}\Big]\Delta t,\nonumber\\
&&N_{n+1}=N_{n}+\Big(\Lambda-d N_{n}\Big)\Delta t,\nonumber
\end{eqnarray}
where $n = 0, 1, \dots, K$. By means of the improved discrete inverse method in subsection \ref{subsection2.1}, we can derive the discrete transmission rate, which is given by
\begin{equation}
\beta_{n}=\max\Bigg\{0,\frac{\Big[[\Lambda-(g+d)S_{n}]\Delta t-(S_{n+1}-S_{n})\Big]N_{n}}{S_{n}I_{n}\Delta t}\Bigg\}, \quad n = 0, 1, \dots, K-1,\label{EQ11}
\end{equation}
and $\beta_{K}=\beta_{K-1}$.

Next, we collect monthly data on scarlet fever from two provinces in China, Jiangxi and Tibet, which exhibited low-prevalence from 2004 to 2019. Assuming that the number of new cases follows a Poisson distribution, we can generate 1000 simulated samples of the case counts for each time point (see Figure \ref{fig3}). The parameters for Model (\ref{EQ9}) are as follows: for Tibet, the recruitment rate of the population, $\Lambda$, is calculated as $2740000 \times \frac{17.40}{1000\times12}$ people/month, where the first term represents the total population of Tibet and the second term denotes the monthly birth rate \cite{TBDNAM2025}. The progression rate from exposed to infected individuals, denoted by $\sigma$, is $30/3$ ${\text{month}}^{-1}$, assuming the latent period of three days \cite{Popular2025}. The recovery rate $\gamma$, which governs the transition from the infected to the recovered state, is $30/7$ $\text{month}^{-1}$. This parameterization assumes an average infectious period of seven days (i.e., $\gamma = 1/7\text{day}^{-1}$) \cite{liu2018resurgence,zhong2020mathematical,zhang2022modeling}. Since scarlet fever cases primarily affect children and adolescents under 15 years old, the maturation rate $g$ is set to $1/(15 \times 12)$ $\text{month}^{-1}$ \cite{WikipediaSF2025}. The natural death rate of the population, $d$, is $1/(75 \times 12)$ ${\rm month}^{-1}$, where 75 represents the average life expectancy \cite{NBSCALE2025}. The initial number of susceptible individuals, $S_0$, is $2740000 \times \frac{653}{2636}$ \cite{PACDRPAM2025}, where the first term denotes the total population of the Tibet Autonomous Region and the second term represents the proportion of the population aged 0-15 years; the initial number of infected individuals, $I_0$, is $\tilde{y}_0$, where $\tilde{y}_0$ is the first data point of the newly reported cases; consequently, the initial number of exposed individuals, $E_0$, is derived as $\tilde{y}_0 / \sigma$; the initial number of recovered individuals, $R_0$, is 0; and the initial number of adult individuals, $A_0$, is $2740000 - 2740000 \times \frac{653}{2636}$ \cite{PACDRPAM2025}. For Jiangxi, the recruitment rate, $\Lambda$, is calculated as $41520000 \times \frac{13.61}{1000 \times 12}$ \cite{TBDNAM2025}. The initial state vector is defined as $(S_0, E_0, I_0, R_0, A_0) = (42840000 \times \frac{8963}{41524}, \tilde{y}_0 / \sigma, \tilde{y}_0, 0, 42840000 - 42840000 \times \frac{8963}{41524})$ \cite{PACDRPAM2025}, corresponding to the initial numbers of susceptible, exposed, infected, recovered, and adult individuals. All other parameter values are consistent with those applied for the Tibet Autonomous Region.

Figure \ref{fig3} illustrates the estimated results of scarlet fever transmission rates from 2004 to 2019 in Jiangxi Province (see Figure \ref{fig3}(A) and (B)) and the Tibet Autonomous Region (see Figure \ref{fig3}(C) and (D)), obtained using our proposed improved discrete inverse method. These results demonstrate the high accuracy of our method, as the estimated new cases (magenta lines in Figure \ref{fig3}(B) and (D)) closely align with the actual new cases in both regions. This is further supported by the narrow 95\% CIs, which largely encompass the true surveillance data points. Moreover, the estimated $\beta(t)$ (blue lines in Figure \ref{fig3}(A) and (C)) effectively captures the dynamic changes in scarlet fever transmission intensity over the study period. Overall, these findings validate that our method achieves high-fidelity fitting to real epidemiological data and reliably reconstructs the time-varying transmission parameters, confirming its robustness in modeling epidemiological dynamics.

\section{Application to multi-strain infectious diseases}\label{section4}
To evaluate the effectiveness of our proposed improved inverse method in reconstructing the time-varying transmission rates of multi-strain infectious diseases, we construct a multi-strain mathematical model where the transmission rates dynamically vary with time. The whole population is divided into four classes: susceptible individuals $(S)$, those exposed with strain $i$ $(E_i)$, those infectious with strain $i$ $(I_i)$, and recovered individuals $(R)$. The total population size is given by
$$N(t)=S(t)+\sum^{n}_{i=1}E_{i}(t)+\sum^{n}_{i=1}I_{i}(t)+R(t).$$
As in model \eqref{EQ1}, $\Lambda$ represents the recruitment rate, and $d$ is the natural mortality rate. The parameter $\delta$ denotes the rate at which recovered individuals lose immunity. The average latent period for strain $i$ is $1/\sigma_{i}$. Furthermore, $\beta_{i}(t)$ and $\gamma_{i}$ represent the time-varying transmission rate and the recovery rate for strain $i$, respectively. The dynamics of our multi-strain model \cite{jing2025infectivity} are described by
\begin{eqnarray}
&&\;\frac{\mbox{d}S(t)}{\mbox{d}t}=\Lambda-\sum^{n}_{i=1}\frac{\beta_{i}(t)S(t)I_{i}(t)}{N(t)}-dS(t)+\delta R(t),\nonumber\\
&&\left.
\begin{split}
&\frac{\mbox{d}E_{i}(t)}{\mbox{d}t}=\frac{\beta_{i}(t)S(t)I_{i}(t)}{N(t)}-\sigma_{i} E_{i}(t)-dE_{i}(t),\qquad i=1,\ldots,n,\\
&\frac{\mbox{d}I_{i}(t)}{\mbox{d}t}=\sigma_{i} E_{i}(t)-\gamma_{i}I_{i}(t)-dI_{i}(t),\qquad\qquad\quad\;\;\; i=1,\ldots,n,\\
\end{split}
\right.\label{EQ12}\\
&&\;\frac{\mbox{d}R(t)}{\mbox{d}t}=\sum^{n}_{i=1}\gamma_{i}I_{i}(t)-dR(t)-\delta R(t).\nonumber
\end{eqnarray}

For each strain $i\in\{1,\dots,n\}$, let $y_{i,j}>0\;(j=0,1,\dots,M,\;i=1,\dots,n)$ denote the observed number of new cases of strain $i$ during day (week, month, etc.) $j$. Through exponential B-spline interpolation, we can obtain smooth and strictly positive approximations of the new cases, denoted as $\tilde{y}_i(t)$. In the Model (\ref{EQ12}), $\sigma_i E_i(t)$ represents the number of new cases of strain $i$ per unit time, thus we set
$$
\sigma_i E_i(t) = \tilde{y}_i(t),\quad t\in[0,M],\; i=1,\dots,n.
$$
Since $\tilde{y}_i(t)$ is differentiable, we have
$$
\frac{{\rm d}E_i(t)}{{\rm d}t}= \frac{1}{\sigma_i}\frac{{\rm d}\tilde{y}_i(t)}{{\rm d}t},\quad t\in[0,M].
$$
We define
$$
H_i(t) := \frac{{\rm d}E_i(t)}{{\rm d}t}, \quad i=1,\dots,n.
$$

Summing all equations of the multi-strain model shows that the total population size
$$
N(t)=S(t)+\sum_{i=1}^{n}\big(E_i(t)+I_i(t)\big)+R(t)
$$
still satisfies
$$
\frac{{\rm d}N(t)}{{\rm d}t}=\Lambda - dN(t),\quad N(0)=N_0,
$$
whose solution is
$$
N(t) = N_0 {\rm e}^{-dt} + \frac{\Lambda}{d}\big(1-{\rm e}^{-dt}\big),\quad t\ge0.
$$

For the infectious classes, each $I_i(t)$ solves the linear non-homogeneous equation
$$
\frac{{\rm d}I_i(t)}{{\rm d}t}
= \sigma_i E_i(t) - (\gamma_i + d) I_i(t)
= \tilde{y}_i(t) - (\gamma_i + d)I_i(t),
$$
so that
$$
I_i(t) = {\rm e}^{-(\gamma_i + d)t}\left(
I_{i0} + \int_0^t {\rm e}^{(\gamma_i + d)s}\tilde{y}_i(s),{\rm d}s
\right),
\quad t\in[0,M],\quad i=1,\dots,n,
$$
where $I_{i0}=I_i(0)$ are the initial conditions.

The recovered class satisfies
$$
\frac{{\rm d}R(t)}{{\rm d}t} = \sum_{i=1}^{n}\gamma_i I_i(t) - (d+\delta)R(t),\quad R(0)=R_0.
$$
Thus,
$$
R(t) = {\rm e}^{-(d+\delta)t}\left(
R_0 + \int_0^t {\rm e}^{(d+\delta)s}\sum_{i=1}^{n}\gamma_i I_i(s){\rm d}s
\right),\quad t\in[0,M].
$$

The susceptible class can be reconstructed from the conservation relation
$$
S(t) = N(t) - \sum_{i=1}^{n}\left(E_i(t)+I_i(t)\right) - R(t),\quad t\in[0,M].
$$

Finally, we use the equations for $E_i(t)$ to recover the transmission rates $\beta_i(t)$. Rearranging the $E_i$ equation gives
$$
\frac{{\rm d}E_i(t)}{{\rm d}t}
= \frac{\beta_i(t)S(t)I_i(t)}{N(t)} - (\sigma_i + d)E_i(t),
$$
which can be solved for $\beta_i(t)$ as
$$
\beta_i(t)
= \frac{\big[H_i(t) + (\sigma_i + d)E_i(t)\big]N(t)}{S(t)I_i(t)},\quad t\in[0,M],\; i=1,\dots,n.
$$
Using $E_i(t)=\tilde{y}_i(t)/\sigma_i$, this becomes
$$
\beta_i(t)
= \frac{\Big[H_i(t) + (\sigma_i + d)\frac{\tilde{y}_i(t)}{\sigma_i}\Big]N(t)}{S(t)I_i(t)},
\quad t\in[0,M],\; i=1,\dots,n.
$$

To enforce the nonnegativity of the transmission rates, we define the inverse estimator of the strain-specific transmission rates as
\begin{equation}
\beta_i(t)= \max\left\{
0,\frac{\Big[H_i(t) + (\sigma_i + d)\frac{\tilde{y}_i(t)}{\sigma_i}\Big]N(t)}{S(t)I_i(t)}\right\},\quad t\in[0,M],\; i=1,\dots,n.
\label{EQ13}
\end{equation}

\begin{figure}[ht]
\centering
{\includegraphics[width=5.5in,height=4in,clip]{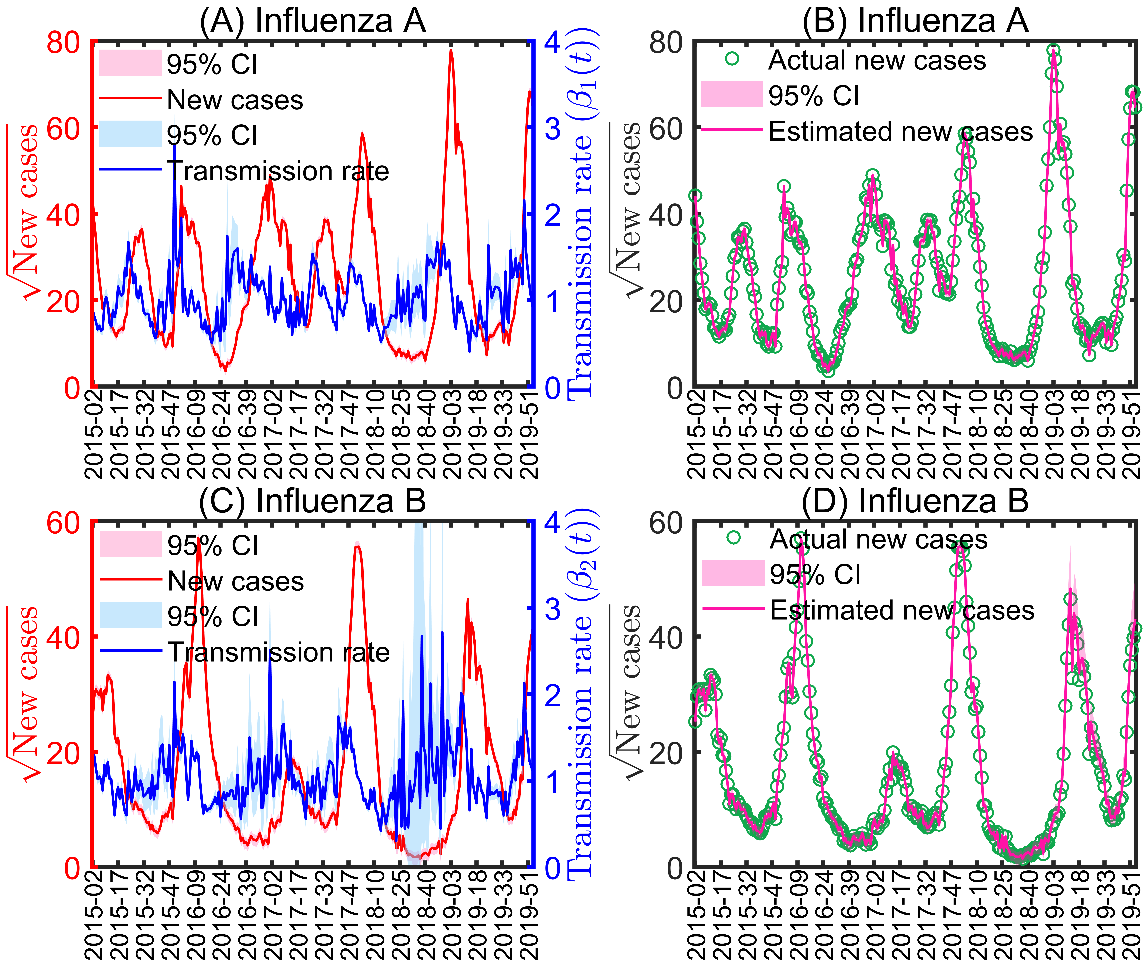} }
\caption{Estimation of transmission rate for influenza A and B in China during the pre-COVID-19 era (week 1 of 2015 to the final week of 2019). Panels (A) and (C) respectively display the square root of the new cases (red curves), transmission rates ($\beta_1(t)$/$\beta_2(t)$) (blue curves), and their 95\% CIs (light shaded areas) for influenza A and B. Panels (B) and (D) show the model fitting results for the square root of actual new cases (green scatter points) of influenza A and B, with pink curves indicating the square root of estimated new cases and pink shaded areas representing the 95\% CIs of the fitted values.}\label{fig4}
\end{figure}

\begin{figure}[ht]
\centering
{\includegraphics[width=5.5in,height=4in,clip]{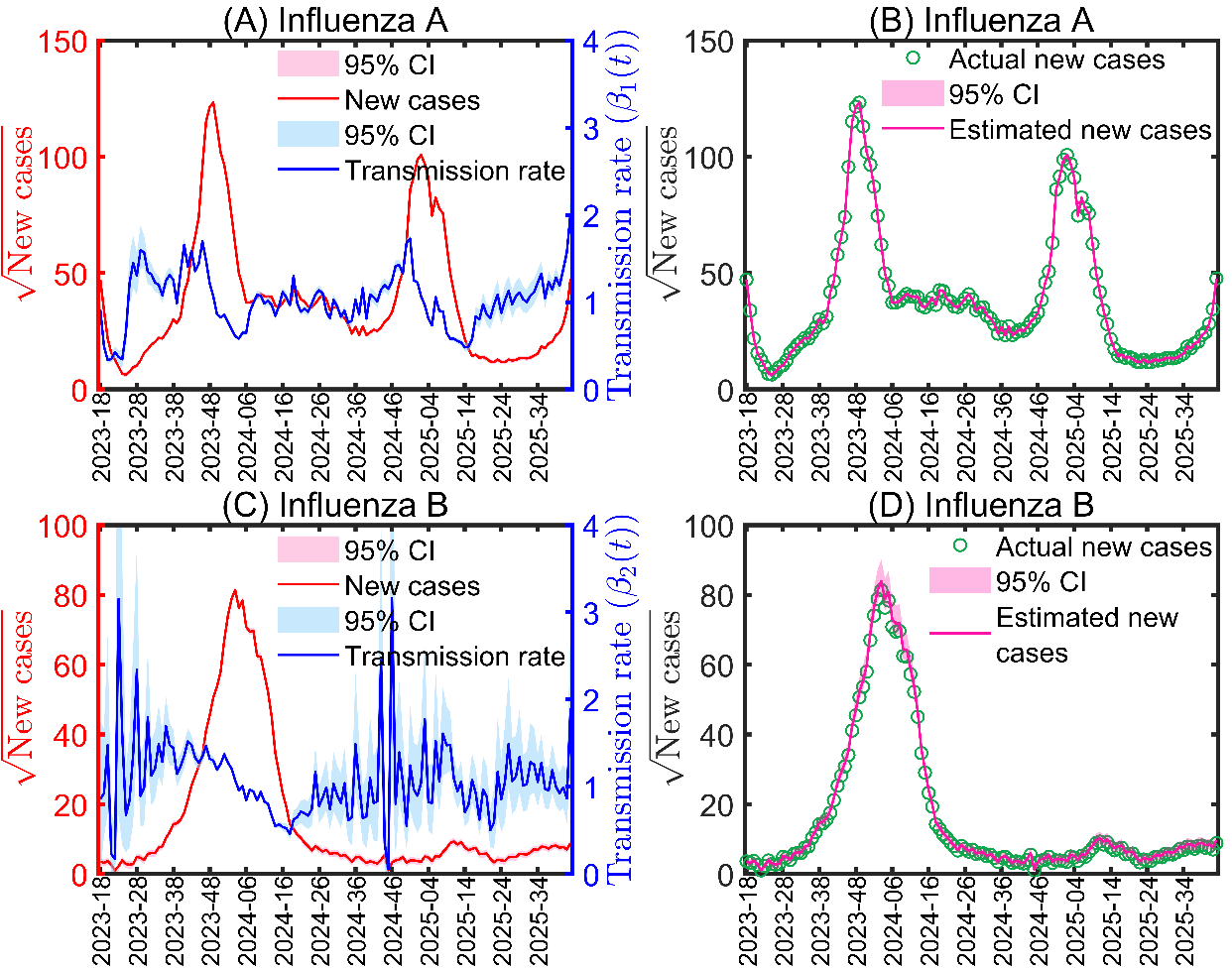} }
\caption{Estimation of transmission rate for influenza A and B in China during the post-COVID-19 era (week 18 of 2023 to week 43 of 2025). Graphic elements in this figure are defined in the same way as those in Figure \ref{fig4}.}\label{fig5}
\end{figure}

We collect weekly confirmed cases of influenza by strain in China from sentinel and non-sentinel surveillance \cite{WHOILSI2025} during the pre-COVID-19 era (week 1 of 2015 to the final week of 2019) (see Figure \ref{fig4}) and during the post-COVID-19 era (week 18 of 2023 to week 43 of 2025) (see Figure \ref{fig5}). As in Section \ref{section3}, we assume that the number of new cases follows a Poisson distribution and generate 1000 simulated samples of case counts at each time point. In the simulations, we assume the initial values for susceptible individuals are set to $S(0) = 0.97 N(0)$. Similarly, the initial values for recovered individuals are given by $R(0) = 0.03N(0)$, where the constant $0.03$ represents the vaccination rate and the initial total population $N(0)$ is 1376460000 (pre-COVID-19) and 1411750000 (post-COVID-19) \cite{NBSCCPSC025}. We also assume that the initial number of individuals infected with strain $i$ is given by $I_{i}(0) = \sigma_{i}E_{i}(0)$, and postulate that this initial number of individuals infected with strain $i$ equals the initial number of new cases, thereby allowing the derivation of the initial number of individuals in the exposed individuals. For $\Lambda$, its value is $1376460000\times\frac{13.83}{1000\times52}$ number/week (pre-COVID-19) and $1411750000\times\frac{6.77}{1000\times52}$ number/week (post-COVID-19) \cite{NBSCBDNF2025,NBSCCPSC025}. The rate of progression from exposed to infected individuals ($\sigma_i$) and the recovery rate ($\gamma_i$) for both strains are set to $7/2\;\text{week}^{-1}$ and $1\;\text{week}^{-1}$, respectively, for both the pre-COVID-19 and post-COVID-19 periods \cite{jing2025infectivity}. $\delta$ maintains a constant $1/52\;\text{week}^{-1}$ across both periods \cite{jing2025infectivity}, while $d$ remains $1/(75 \times 52)\;\text{week}^{-1} $ in both stages \cite{NBSCALE2025}.

\begin{table}[!ht]
\centering
\caption{The average transmission rates of different influenza strains.}
\begin{tabular}{|c|c|c|c|c|c|c|}
\hline
\multirow{2}{*}{\diagbox{Strain}{Time}} & \multicolumn{4}{|c|}{pre-COVID-19} & \multicolumn{2}{|c|}{post-COVID-19}\\
\cline{2-7}
         &2015-2016&2016-2017&2017-2018&2018-2019& 2023-2024 & 2024-2025\\
\hline
      Influenza A & 1.0458 &1.0277&0.9476&1.1097& 1.0571 & 0.9881\\
\hline
      Influenza B & 1.1582 &1.1507&1.0203&1.3084& 1.0665 & 1.1289\\
\hline
\end{tabular}\label{Table3}
\end{table}

To characterize the dynamic transmission of influenza, Figures \ref{fig4}, \ref{fig5}, and Table \ref{Table3} compare the time-varying new cases, time-varying transmission rates, and the average transmission rate during the influenza peak period (from week 40 of the previous year to week 17 of the following year) before and after the COVID-19 pandemic.

Figure \ref{fig4} shows the pre-COVID-19 era (week 1 of 2015 to the final week of 2019). For influenza A (Panel (A)) and influenza B (Panel (C)), the weekly new cases (red curve, with light pink shading indicating the 95\% CI) exhibited moderate periodic fluctuations (peak values below 6,000 for influenza A and below 4,000 for influenza B). The blue curves (representing the transmission rate for influenza A and B, respectively) and their light blue 95\% CIs closely tracked these trends. Furthermore, Panels (B) and (D) confirm validity of the model, as the estimated new cases (pink curve, with 95\% CI) align closely with the actual new cases (green scatter points). In contrast, Figure \ref{fig5} illustrates the dynamics of markedly intensified transmission in the post-COVID-19 era (week 18 of 2023 to week 43 of 2025). During this interval, the weekly numbers of new cases of influenza A and influenza B rise to approximately 15,000 and 7,000, respectively, far exceeding their pre-pandemic peaks, and both strains display substantially larger fluctuations. Despite the increased intensity of transmission, the estimated new cases (pink curve) consistently matches the actual data (green points), underscoring the robustness of the proposed inverse method.

Before the COVID-19 pandemic, the transmission rate of influenza B (1.0203-1.3084) consistently exceeded that of influenza A (0.9476-1.1097) (see Table \ref{Table3}). In the post-pandemic period, the transmission rate of influenza A (1.0571 in the 2023-2024 season and 0.9881 in the 2024-2025 season) remains comparable to its pre-pandemic levels, whereas the incidence of influenza B rebounds to 1.1289 in the 2024-2025 season, approaching its pre-pandemic peak of 1.3084 in 2018-2019 (see Table \ref{Table3}). This pattern suggests a gradual return of influenza B transmission intensity toward its pre-COVID-19 level.

\section{Application to age-structured infectious diseases}\label{section5}
In this section, we numerically test the robustness of our proposed inverse method for reconstructing the transmission rates in an age-structured model. For this purpose, the total population is divided into $n$ age groups \cite{jing2025infectivity}. Each group is further stratified into four compartments: susceptible ($S_k$), exposed ($E_k$), infectious ($I_k$), and recovered ($R_k$) individuals. The total population size of the $k$-th age group at time $t$ is given by $N_k(t) = S_k(t) + E_k(t) + I_k(t) + R_k(t)$.  The overall total population, $N(t)$, is obtained by summing $N_k(t)$ over all age groups. The population dynamics of the $k$-th age group are governed by the following processes: the total population grows at a constant rate $\Lambda$, with newborn babies typically assumed to enter the susceptible compartment of the first age group, while the $k$-th age group experiences natural mortality at a rate $d_k N_k(t)$ and ages into the next age group (i.e., group $k+1$) at a rate $\alpha_k N_k(t)$. Here, $d_k$ represents the natural mortality rate and $\alpha_k$ denotes the transfer rate out of the $k$-th age group.

The forces or hazard rates of infection among susceptible individuals \cite{feng2020influence} in age-group $k$ are
\begin{equation*}
\lambda_{k}(t)=\beta_{k}(t)\sum_{j = 1}^{n}\frac{C_{kj}I_{j}(t)}{N_{j}(t)}, \quad \text{for}\quad k=1,\ldots,n,
\end{equation*}
where the time-varying parameter $\beta_{k}(t)$ denotes the probability of infection for susceptible individuals in age group $k$ upon contact with an infectious individual, $C_{kj}$ represents the per-unit-time contact frequency between individuals in age group $k$ and those in age group $j$, while $\frac{I_{j}}{N_{j}}$ indicates the probability of encountering an infectious individual from age-group $j$.

For the $k$-th age group, susceptible individuals become infected with the influenza virus and move to the exposed class at a rate $\lambda_k(t)S_k(t)$. Exposed individuals progress to the infectious class at a rate $\sigma_k E_k(t)$, where $1/\sigma_k$ represents the duration of the latent period. Infectious individuals transition to the recovered class at a rate $\gamma_k I_k(t)$, with $1/\gamma_k$ denoting the duration of the infectious period. Recovered individuals do not acquire lifelong immunity and transition to the susceptible class at a rate $\delta_k R_k(t)$, where $1/\delta_k$ represents the duration of immune protection. Based on these mechanisms, our model is given as follows by
\begin{eqnarray}
&&\;\frac{\mbox{d}S_{1}(t)}{\mbox{d}t}=\Lambda-\lambda_{1}(t)S_{1}(t)-d_{1}S_{1}(t)-\alpha_{1}S_{1}(t)+\delta_{1} R_{1}(t),\nonumber\\
&&\;\frac{\mbox{d}S_{k}(t)}{\mbox{d}t}=\alpha_{k-1}S_{k-1}(t)-\lambda_{k}(t)S_{k}(t)-d_{k}S_{k}(t)-\alpha_{k}S_{k}(t)+\delta_{k} R_{k}(t),\nonumber\\
&&\;\frac{\mbox{d}E_{1}(t)}{\mbox{d}t}=\lambda_{1}(t)S_{1}(t)-\sigma_{1} E_{1}(t)-d_{1}E_{1}(t)-\alpha_{1}E_{1}(t),\nonumber\\
&&\left.
\begin{split}
&\frac{\mbox{d}E_{k}(t)}{\mbox{d}t}=\alpha_{k-1}E_{k-1}(t)+\lambda_{k}(t)S_{k}(t)-\sigma_{k}E_{k}(t)-d_{k}E_{k}(t)-\alpha_{k}E_{k}(t),\\
&\frac{\mbox{d}I_{1}(t)}{\mbox{d}t}=\sigma_{1} E_{1}(t)-\gamma_{1}I_{1}(t)-d_{1}I_{1}(t)-\alpha_{1}I_{1}(t),\\
\end{split}
\right.\label{EQ14}\\
&&\;\frac{\mbox{d}I_{k}(t)}{\mbox{d}t}=\alpha_{k-1}I_{k-1}(t)+\sigma_{k} E_{k}(t)-\gamma_{k}I_{k}(t)-d_{k}I_{k}(t)-\alpha_{k}I_{k}(t),\nonumber\\
&&\;\frac{\mbox{d}R_{1}(t)}{\mbox{d}t}=\gamma_{1}I_{1}(t)-d_{1}R_{1}(t)-\alpha_{1}R_{1}(t)-\delta_{1} R_{1}(t),\nonumber\\
&&\;\frac{\mbox{d}R_{k}(t)}{\mbox{d}t}=\alpha_{k-1} R_{k-1}(t)+\gamma_{k}I_{k}(t)-d_{k}R_{k}(t)-\alpha_{k}R_{k}(t)-\delta_{k} R_{k}(t), \nonumber
\end{eqnarray}
where $k=2,\ldots,n$.

To apply our inverse methodology to the age-structured model (\ref{EQ14}), we let $y_{k,j}>0$ denote the number of newly observed cases in age group $k$ on day (week, month, etc.) $j$, for $k=1,\dots,n$ and $j=0,1,\dots,M$. Similar to Model (\ref{EQ1}), we assume that these observations correspond to the transition rate $\sigma_k E_k(t)$ from the latent to the infectious compartment in age group $k$. We obtain a smooth and strictly positive approximation for new cases in the $k$-th age group via exponential B-spline interpolation, denoted as $\tilde{y}_{k}(t)$. Since $\sigma_k E_k(t)$ represents the number of new cases per unit time in age group $k$, we have
$$
\sigma_k E_k(t) = \tilde{y}_k(t),\quad t\in[0,M],\quad k=1,\dots,n.
$$
Consequently, we define the derivative of $E_k(t)$ is
$$
H_k(t):=\frac{{\rm d}E_k(t)}{{\rm d}t}
= \frac{1}{\sigma_k}\frac{{\rm d}\tilde{y}_k(t)}{{\rm d}t},\quad t\in[0,M],\quad k=1,\dots,n.
$$

Next, we solve for $N_k(t)$, $I_k(t)$, and $R_k(t)$ for each age group. To do this, we note that the total population $N_k(t)=S_k(t)+E_k(t)+I_k(t)+R_k(t)$ of the $k$-th age group in Model (\ref{EQ14}) satisfies a linear system
\begin{eqnarray}
&&\frac{{\rm d}N_1(t)}{{\rm d}t}=\Lambda-(d_1+\alpha_1)N_1(t),\nonumber\\
&&\frac{{\rm d}N_k(t)}{{\rm d}t}=\alpha_{k-1}N_{k-1}(t)-(d_k+\alpha_k)N_k(t),\quad k=2,\ldots,n.\nonumber
\end{eqnarray}
The corresponding solutions are
\begin{eqnarray}
&&N_1(t)=N_1(0){\rm e}^{-(d_1+\alpha_1)t}+\Lambda\int_0^t {\rm e}^{-(d_1+\alpha_1)(t-s)}{\rm d}s,\nonumber\\
&&N_k(t)=N_k(0){\rm e}^{-(d_k+\alpha_k)t}+\alpha_{k-1}\int_0^t {\rm e}^{-(d_k+\alpha_k)(t-s)}N_{k-1}(s){\rm d}s,\quad k=2,\ldots,n.\nonumber
\end{eqnarray}
Similarly, the infection compartment ($I_k(t)$) and the recovery compartment ($R_k(t)$) satisfy linear non-homogeneous equations, which depend on the number of exposed compartment ($E_k(t)$). For the first age group ($k=1$), we have
\begin{eqnarray}
&&\frac{{\rm d}I_1(t)}{{\rm d}t}=\sigma_1E_1(t)-(\gamma_1+d_1+\alpha_1)I_1(t),\nonumber\\
&&\frac{{\rm d}R_1(t)}{{\rm d}t}=\gamma_1I_1(t)-(d_1+\alpha_1+\delta_1)R_1(t),\nonumber
\end{eqnarray}
which lead to
\begin{eqnarray}
&&I_1(t)={\rm e}^{-(\gamma_1+d_1+\alpha_1)t}\left(I_1(0)+\int_0^t {\rm e}^{(\gamma_1+d_1+\alpha_1)s}\sigma_1E_1(s){\rm d}s\right),\nonumber\\
&&R_1(t)={\rm e}^{-(d_1+\alpha_1+\delta_1)t}\left(R_1(0)+\int_0^t {\rm e}^{(d_1+\alpha_1+\delta_1)s}\gamma_1I_1(s){\rm d}s\right).\nonumber
\end{eqnarray}
For $2\leq k\leq n$, we obtain
\begin{eqnarray}
&&\frac{{\rm d}I_k(t)}{{\rm d}t}=\alpha_{k-1}I_{k-1}(t)+\sigma_kE_k(t)-(\gamma_k+d_k+\alpha_k)I_k(t),\nonumber\\
&&\frac{{\rm d}R_k(t)}{{\rm d}t}=\alpha_{k-1}R_{k-1}(t)+\gamma_kI_k(t)-(d_k+\alpha_k+\delta_k)R_k(t),\nonumber
\end{eqnarray}
which lead to
\begin{eqnarray}
&&I_k(t)={\rm e}^{-(\gamma_k+d_k+\alpha_k)t}\left(
I_k(0)+\int_0^t {\rm e}^{(\gamma_k+d_k+\alpha_k)s}\big(\alpha_{k-1}I_{k-1}(s)+\sigma_kE_k(s)\big){\rm d}s\right),\nonumber\\
&&R_k(t)={\rm e}^{-(d_k+\alpha_k+\delta_k)t}\left(
R_k(0)+\int_0^t {\rm e}^{(d_k+\alpha_k+\delta_k)s}\big(\alpha_{k-1}R_{k-1}(s)+\gamma_kI_k(s)\big){\rm d}s\right).\nonumber
\end{eqnarray}
Once $N_k(t)$, $E_k(t)$, $I_k(t)$, and $R_k(t)$ are known, the number of susceptible individuals in the $k$-th age group is given by
$$
S_k(t)=N_k(t)-E_k(t)-I_k(t)-R_k(t),\quad k=1,\dots,n.
$$

Combining these identities with the definition of $\lambda_k(t)$, we obtain
\begin{equation*}
\beta_k(t)
=\frac{\lambda_k(t)}{\sum_{j=1}^n \frac{C_{kj}I_j(t)}{N_j(t)}}
=\frac{B_k(t)}{S_k(t)\sum_{j=1}^n \frac{C_{kj}I_j(t)}{N_j(t)}},
\quad t\in[0,M],
\end{equation*}
where
\begin{eqnarray}
&&B_1(t):=H_1(t)+(\sigma_1+d_1+\alpha_1)\frac{\tilde{y}_1(t)}{\sigma_1},\nonumber\\
&&B_k(t):=H_k(t)-\alpha_{k-1}\frac{\tilde{y}_{k-1}(t)}{\sigma_{k-1}}
+(\sigma_k+d_k+\alpha_k)\frac{\tilde{y}_k(t)}{\sigma_k},\quad k=2,\ldots,n.\nonumber
\end{eqnarray}
Finally, to enforce the biological constraint $\beta_k(t)\geq 0$, we define
\begin{equation}
\beta_k(t)
=\max\left\{0,\frac{B_k(t)}{S_k(t)\sum_{j=1}^n \frac{C_{kj}I_j(t)}{N_j(t)}}
\right\},\quad t\in[0,M],\quad k=1,\dots,n.\label{EQ15}
\end{equation}
\begin{figure}[ht]
\centering
{\includegraphics[width=4in,height=3.5in,clip]{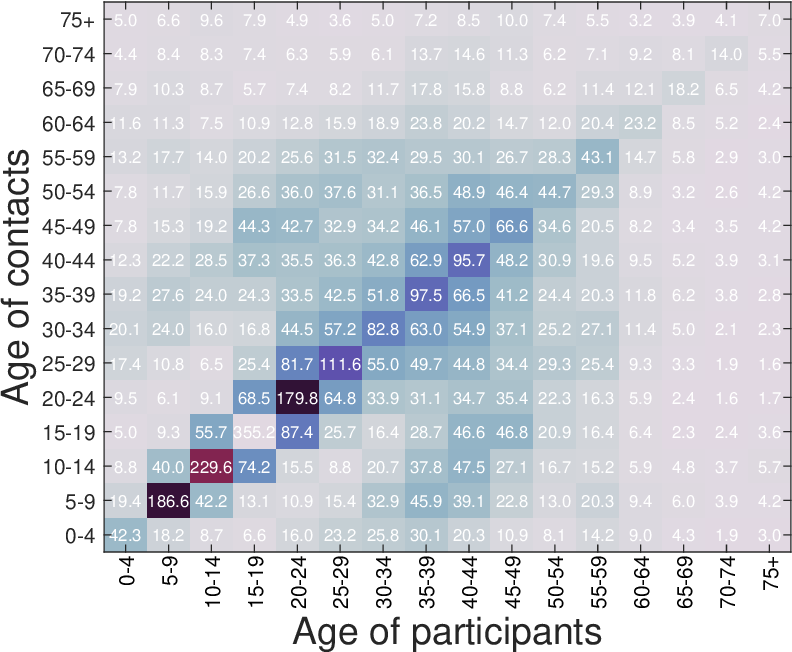} }
\caption{The monthly average number of contacts per person in the participant age-group.}\label{fig6}
\end{figure}
\begin{figure}[ht]
\centering
{\includegraphics[width=5.5in,height=4in,clip]{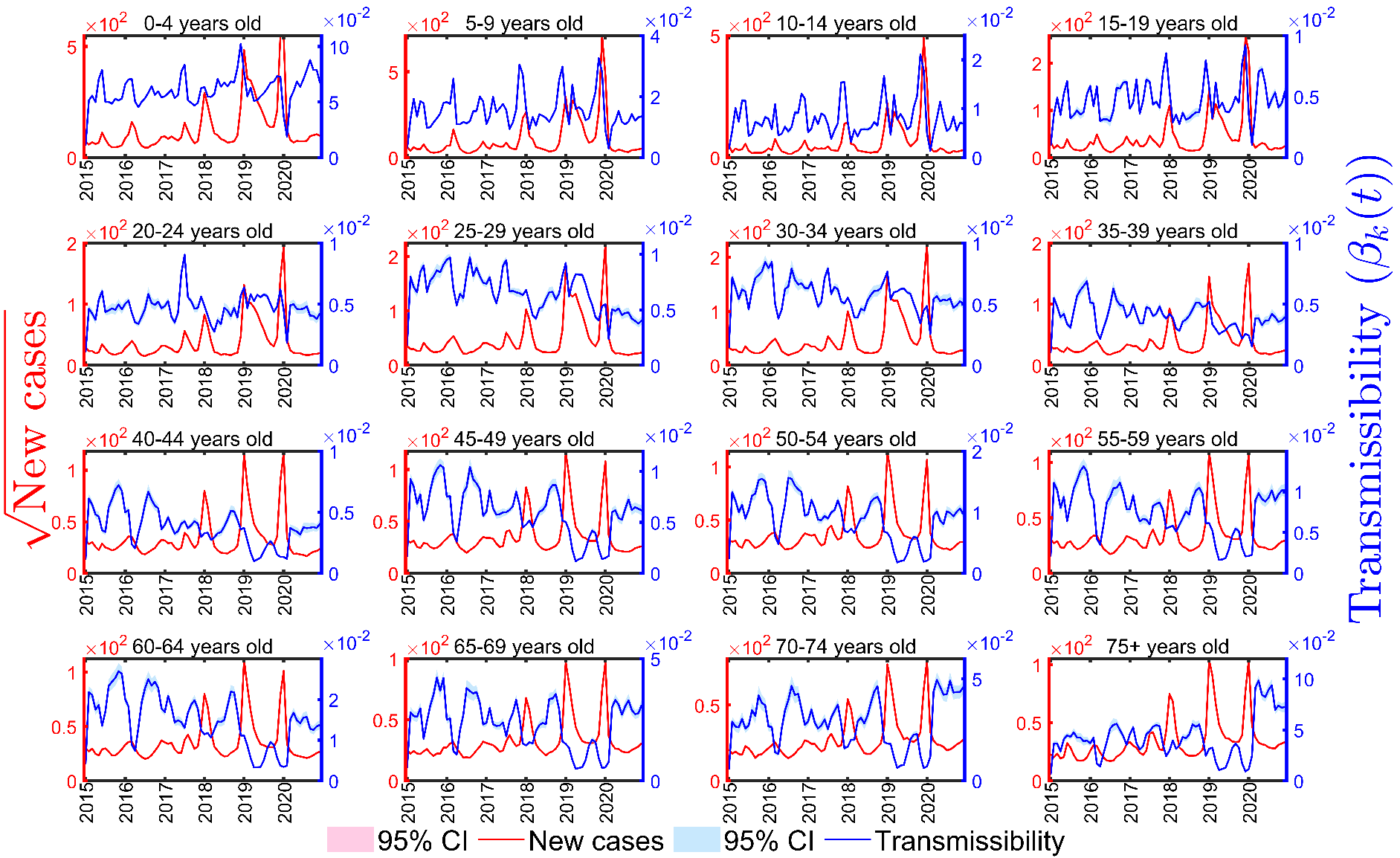} }
\caption{Temporal variations of the square root of the new cases and transmissibility across distinct age groups (0-4 years, 5-9 years, 10-14 years, 15-19 years, 20-24 years, 25-29 years, 30-34 years, 35-39 years, 40-44 years, 45-49 years, 50-54 years, 55-59 years, 60-64 years, 65-69 years, 70-74 years, and 75+ years) between 2015 and 2020. In each subfigure, the red line denotes the square root of the new cases (accompanied by the 95\% CI shaded in pink), and the blue line represents transmissibility (with the 95\% CI shaded in light blue).}\label{fig7}
\end{figure}
\begin{figure}[ht]
\centering
{\includegraphics[width=5.5in,height=4in,clip]{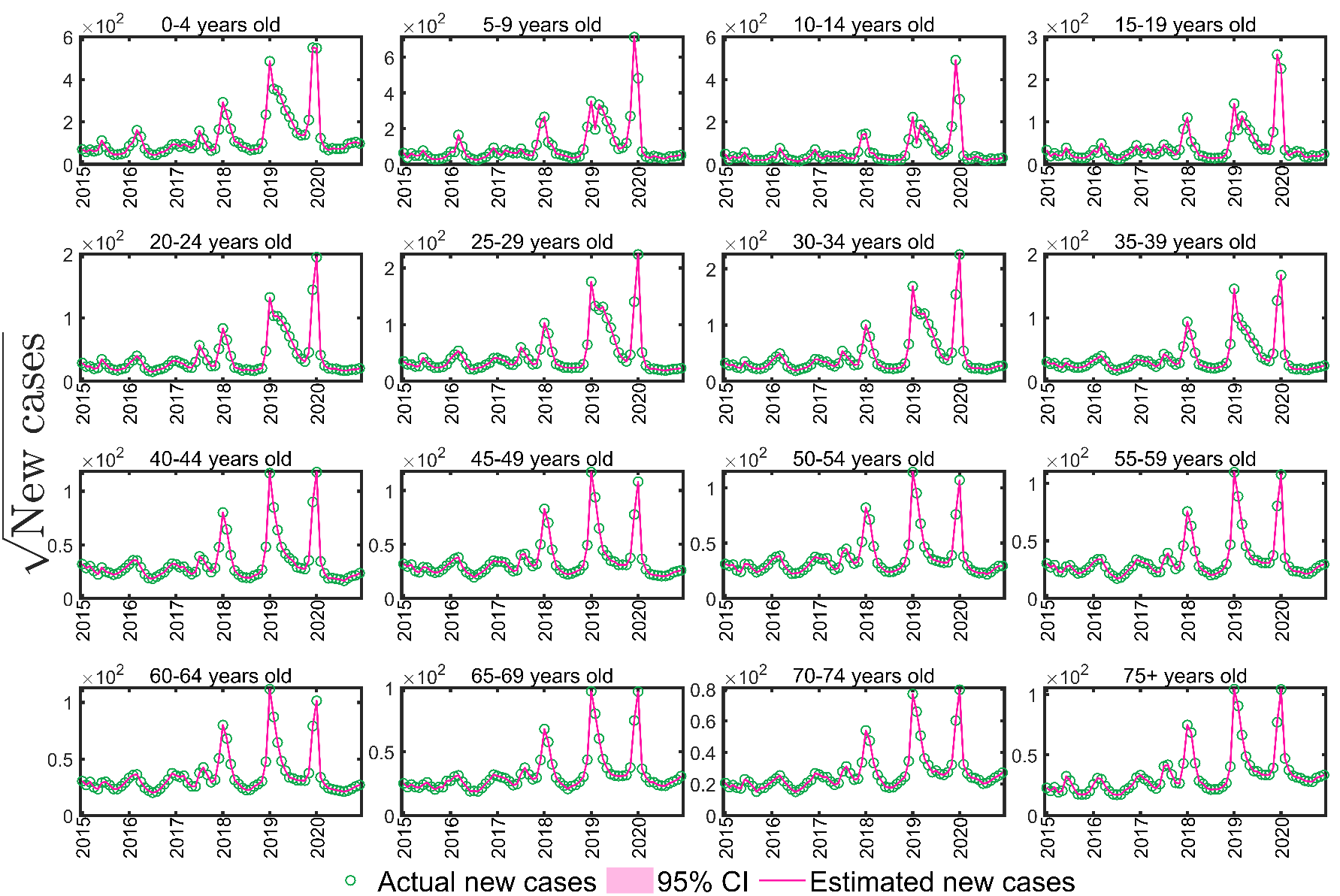} }
\caption{Temporal trends of the square root of the actual and estimated new cases across distinct age groups (0-4 years, 5-9 years, 10-14 years, 15-19 years, 20-24 years, 25-29 years, 30-34 years, 35-39 years, 40-44 years, 45-49 years, 50-54 years, 55-59 years, 60-64 years, 65-69 years, 70-74 years, and 75+ years) from 2015 to 2020. In each subfigure, green circles represent the square root of the actual new cases, the pink solid line denotes the square root of the estimated new cases, and the pink shaded area indicates the 95\% CI for the estimated new cases.}\label{fig8}
\end{figure}

In the simulation, the population is divided into 16 age groups: 0-4 years old, 5-9 years old, 10-14 years old, 15-19 years old, 20-24 years old, 25-29 years old, 30-34 years old, 35-39 years old, 40-44 years old, 45-49 years old, 50-54 years old, 55-59 years old, 60-64 years old, 65-69 years old, 70-74 years old, and 75+ years old. Meanwhile, we collect the monthly number of newly confirmed influenza cases across 16 age groups in mainland China \cite{DataDirectory25} (see Figure \ref{fig7}). Using the assumption of a Poisson distribution for new cases, 1000 simulated case count samples are generated for each time point.
Subsequently, we set the necessary parameters for the simulation. For the initial values in Model (\ref{EQ14}), we define
$$
S_k(0) = 0.97N_k(0),\;E_{k}(0)=\frac{I_{k}(0)}{\sigma_{k}},\;I_{k}(0)=\tilde{y}_k(0), \; R_k(0) = 0.03N_k(0),\; k = 1, \dots, 16,
$$
where 0.03 represents the vaccination rate and the population sizes \(N_k(0)\) for each age group are given by
\begin{eqnarray}
&&N_1(0) = 75{,}532{,}610, \; N_2(0) = 70{,}881{,}549, \; N_3(0) = 74{,}908{,}462, \; N_4(0) = 99{,}889{,}114, \nonumber\\
&&N_5(0) = 127{,}412{,}518, \; N_6(0) = 101{,}013{,}852, \; N_7(0) = 97{,}138{,}203, \; N_8(0) = 118{,}025{,}959, \nonumber\\
&&N_9(0) = 124{,}753{,}964, \; N_{10}(0) = 105{,}594{,}553, \; N_{11}(0) = 78{,}753{,}171, \; N_{12}(0) = 81{,}312{,}474, \nonumber\\
&&N_{13}(0) = 58{,}667{,}282, \; N_{14}(0) = 41{,}113{,}282, \; N_{15}(0) = 32{,}972{,}397, \; N_{16}(0) = 44{,}841{,}479.\nonumber
\end{eqnarray}
We assume the average life expectancy in China to be 80 years. Accordingly, the monthly natural mortality rates for each age group are set as follows
\begin{eqnarray}
&&d_1= 1/(80\times12),\; d_2= 1/(75\times12),\; d_3= 1/(70\times12),\; d_4= 1/(65\times12),\nonumber\\
&&d_5= 1/(60\times12),\; d_6= 1/(55\times12),\; d_7= 1/(50\times12),\; d_8= 1/(45\times12),\nonumber\\
&&d_9= 1/(40\times12),\; d_{10}= 1/(35\times12),\; d_{11}= 1/(30\times12),\; d_{12}= 1/(25\times12),\nonumber\\
&&d_{13}= 1/(20\times12),\; d_{14}= 1/(15\times12),\; d_{15}= 1/(10\times12),\; d_{16}= 1/(5\times12).\nonumber
\end{eqnarray}
The aging rates across age groups are given by $\alpha_{1}=\alpha_{2}=\cdots=\alpha_{15}=1/(5\times12)\;{\rm month}^{-1}$ and $\alpha_{16}=0\;{\rm month}^{-1}$. The recruitment rate for the first age group is calculated as $\Lambda=1374620000\times\frac{12.07}{1000\times12}\;{\rm number/month}$,
where $\frac{12.07}{1000}$ denotes the annual birth rate per 1000 population.
For the rate of progression from exposed to infected individuals, recovery rate, and immunity loss rate, we assume identical values across all age groups:
$\sigma_1=\cdots=\sigma_{16}=30/2\;{\rm month}^{-1}$, $\gamma_1=\cdots=\gamma_{16}=30/7\;{\rm month}^{-1}$, and $\delta_1=\cdots=\delta_{16}=1/12\;{\rm month}^{-1}$. The contact rates for the 16 age groups are obtained from Reference \cite{prem2017projecting} (see Figure \ref{fig6}).

Figure \ref{fig7} shows the temporal variations in new cases and transmissibility across 16 age groups (0-4 years old to 75+ years old) from 2015 to 2020.
In each subfigure, the red line (with pink shading representing the 95\% CI) represents the time series of new cases, while the blue line (with light blue shading representing the 95\% CI) indicates transmissibility. Temporally, both new cases and transmissibility exhibit marked fluctuations throughout the study period, and distinctly asynchronous temporal dynamics can be observed between new cases and transmissibility across all age groups.

Figure \ref{fig8} presents the temporal variations of actual and estimated new cases across 16 age groups from 2015 to 2020. Here, the green circles denote the actual new cases, the pink solid line represents the estimated new cases, and the pink shaded area corresponds to the 95\% CI of the estimates.
A key finding is that the majority of the actual new cases (green circles) fall within the 95\% CI of the estimated new cases across all age groups and time points. This alignment indicates that the model effectively captures the temporal dynamics of new case incidence, generating reliable and well-calibrated estimates for each age group.

\section{Discussion}\label{section6}
In this work, we developed and validated an improved inverse method for estimating time-varying transmission rates in infectious disease models. The key innovation of this work lies in the application of exponential B-spline interpolation, which addresses challenges posed by low-prevalence diseases where conventional data interpolation techniques often lead to unrealistic negative values. By preserving non-negativity in transmission rates and providing smooth estimates, our approach overcomes limitations of previous methods, proving particularly valuable in scenarios such as early epidemic phases, seasonal outbreaks during low-prevalence periods, and post-intervention surveillance.

We tested our method on a series of epidemiological models, including a general non-autonomous SEIR model, a data-driven childhood infectious disease model for scarlet fever in Jiangxi Province and the Tibet Autonomous Region, a multi-strain influenza model, and an age-structured influenza model. The results demonstrate that our approach outperforms existing techniques, particularly in terms of accuracy and computational efficiency. For instance, when applied to our assumed infectious disease data, the estimated transmission rates closely align with the observed trends, even in scenarios with low-prevalence data. This highlights the robustness of the method in handling datasets that would otherwise lead to biased or unreliable estimates.

We applied the improved inverse method to a scarlet fever transmission model with low-prevalence, successfully estimating accurate time-varying transmission rates. The method also demonstrated its broad applicability when applied to a multi-strain influenza model with time-varying transmission dynamics. Our research results, particularly regarding influenza transmission dynamics in China before and during the COVID-19 pandemic, show that the method effectively captures temporal fluctuations in transmission, providing crucial insights into the evolution of epidemiological patterns before and after the pandemic. Furthermore, applying the method to an age-structured influenza transmission model confirmed its capability to handle complex models featuring multiple compartments and varying transmission dynamics across different age groups.

One of the primary advantages of this improved inverse method is its computational efficiency. Unlike techniques based on machine learning or probabilistic inference, which are often computationally intensive, the continuous and discrete inverse approaches adopted here rely on straightforward mathematical derivations, making them faster and more suitable for real-time public health applications. Furthermore, exponential B-spline interpolation ensures that the results remain biologically plausible and computationally stable, even in the presence of noisy or sparse data.

Despite these strengths, several limitations should be acknowledged. First, although the method performs well in low-incidence settings, its effectiveness may diminish if the underlying model assumptions do not accurately reflect the dynamics of the disease under study. Second, the use of exponential B-spline interpolation requires that newly observed cases satisfy $y_j > 0$ for $j = 0, 1, \dots, M$. In real-world low-prevalence scenarios, zero cases $(y_j = 0)$ may occur, necessitating the addition of small positive noise to handle such instances. However, this adjustment does not compromise the reconstruction of time-varying transmission rates. Finally, while the method effectively estimates transmission rates for homogeneous, multi-strain, and discrete age-structured models, extending it to continuous age-structured models or spatially heterogeneous frameworks, such as those described by reaction-diffusion equations, may require further refinement.

In summary, the improved inverse method offers distinct advantages for estimating time-varying transmission rates in infectious disease models. Its applicability to low-prevalence diseases, multi-strain models, and age-structured frameworks makes it a valuable tool for epidemiological research and public health decision-making. Our future work could focus on adapting the method to more complex systems and exploring its integration with other data sources—such as genomic surveillance or social contact patterns—to further enhance its accuracy and robustness.

\section*{Acknowledgments}
We would like to express our gratitude for the assistance provided by ChatGPT in polishing the writing.

\bibliographystyle{elsarticle-num}
\bibliography{JSL}
\end{document}